\documentclass[letterpaper]{article}
\usepackage{aaai20}
\usepackage{times}
\usepackage{helvet}
\usepackage{courier}
\usepackage{amsmath}
\usepackage{tikz}
\usepackage{numprint}
\usepackage{multirow}
\usepackage{etoolbox,siunitx}
\usepackage{graphicx}
\robustify\bfseries
\usepackage{caption}
\usepackage{subcaption}
\usepackage{tcolorbox}
\usepackage{xcolor}
\usepackage{bbm}
\usepackage{slashbox}

\usetikzlibrary{fit,positioning,arrows,automata,calc}
\tikzset{
  main/.style={circle, minimum size = 5mm, thick, draw=black!80, node distance = 10mm},
  connect/.style={-latex, thick},
  box/.style={rectangle, draw=black!100}
}
\begin{document}
\title{Trawling for Trolling: A Dataset}

%\author{Anonymous Submission}
\author{Hitkul\textsuperscript{\rm 1}\thanks{Equal contribution by the authors.}, Karmanya Aggarwal\textsuperscript{\rm 1}\footnotemark[1], Pakhi Bamdev\textsuperscript{\rm 1}\footnotemark[1]\\ \Large \textbf{Debanjan Mahata\textsuperscript{\rm 2}\thanks{Author participated in this research as an Adjunct Faculty at IIIT-Delhi.}, Rajiv Ratn Shah\textsuperscript{\rm 1}, Ponnurangam Kumaraguru\textsuperscript{\rm 1}}
\\ % All authors must be in the same font size and format. Use \Large and \textbf to achieve this result when breaking a line
% \textsuperscript{\rm 1}\thanks{equal contribution by the authors}
\ \textsuperscript{\rm 1}IIIT-Delhi, India, \textsuperscript{\rm 2}Bloomberg, New York\\ %If you have multiple authors and multiple affiliations
% use superscripts in text and roman font to identify them. For example, Sunil Issar,\textsuperscript{\rm 2} J. Scott Penberthy\textsuperscript{\rm 3} George Ferguson,\textsuperscript{\rm 4} Hans Guesgen\textsuperscript{\rm 5}. Note that the comma should be placed BEFORE the superscript for optimum readability
% Okhla Industrial Estate, Phase III\\
% New Delhi, Delhi 110020\\
(hitkuli, karmanyaa, pakhii, rajivratn, pk)@iiitd.ac.in, dmahata@bloomberg.net % email address must be in roman text type, not monospace or sans serif
}

\maketitle
\begin{abstract}
\begin{quote}
The ability to accurately detect and filter offensive content automatically is important to ensure a rich and diverse digital discourse. Trolling is a type of hurtful or offensive content that is prevalent in social media, but is underrepresented in datasets for offensive content detection. In this work, we present a dataset that models trolling as a subcategory of offensive content. The dataset was created by collecting samples from well-known datasets and reannotating them along precise definitions of different categories of offensive content. The dataset has 12,490 samples, split across 5 classes; Normal, Profanity, Trolling, Derogatory and Hate Speech. It encompasses content from Twitter, Reddit and Wikipedia Talk Pages. Models trained on our dataset show appreciable performance without any significant hyperparameter tuning and can potentially learn meaningful linguistic information effectively. We find that these models are sensitive to data ablation which suggests that the dataset is largely devoid of spurious statistical artefacts that could otherwise distract and confuse classification models.\footnote{This paper contains words and phrases which are considered highly offensive. They do not reflect the views of any of the authors and are intended to be purely demonstrative.}

\end{quote}
\end{abstract}

\section{Introduction}
\label{introduction_section}

\textit{``Fuck you, Smith. Please have me notified when you die. I want to dance on your grave".} Superficially, this message expresses glee at a person's death and thus could cause distress. However, the undercurrent of humour in the message is undeniable - it's highly likely that the message is figurative and is meant to be perhaps slightly mean. In essence, this message constitutes Trolling\footnote{Trolling is defined as content that is intended to be disruptive, inflammatory or mild-mannered insults} - a type of social interaction on the internet that is widespread enough to have entered the general lexicon. Our analysis of popular, publicly available datasets of offensive content \cite{zampieri2019semeval,davidson2017automated,founta2018large,salminen2018anatomy} reveals that they largely ignore the existence of trolling; utilizing labelling schemes that are either too coarse to properly distinguish between the different kinds of offensive content; or too finely focused on subcategories of hate speech \cite{elsherief2018hate}. Founta et al. is the exception, however, it also ignores the existence of trolling content \cite{founta2018large}.

Content that is hateful, oppressive, insulting and obscene can have far-flung repercussions, particularly when amplified through an echo chamber of isolation \cite{doi:10.1177/0956797615594620}. Organizations that own these social media platforms are thus engaged in a balancing act between curtailing free speech and removing bad actors. This makes the identification and filtering of offensive content from social media critically important. To do this, most platforms currently employ some sort of algorithmic filtering backed by manual review. However, the volume of data generated, in conjunction with diverse user demographics makes this task very difficult\footnote{https://www.forbes.com/sites/fruzsinaeordogh/2019/03/11/\\twitters-anti-harassment-tools-reviewed/\#739f65d91e13}. People interact in a variety of ways; similar phrases can mean drastically different things depending upon the cultural and societal context. Finally, social media websites pose a challenge for automated filtering and nlp techniques due to their idiosyncratic language, unusual structure and ambiguous representation of discourse. Information extraction methods also often give poor results when applied in such settings \cite{ritter2011named}.

Though a substantial effort has been made to solve offensive content detection\cite{Fortuna:2018:SAD:3236632.3232676}; terms like \emph{trolling}, \emph{hate speech}, \emph{profanity} and \emph{cyberbullying} are often overloaded or used interchangeably, causing ambiguity and limiting the efficacy of classification models. Ambiguity translates to problematic scenarios in the real world; groups or individuals that indulge in trolling behaviour are occasionally considered proponents of hate speech and de-platformed. This sort of inadvertent censorship has negative repercussions --- not only does it help shape a narrow, intolerant public discourse, but also these censored groups and individuals often develop a feeling of persecution and alienation \cite{doi:10.1177/0267323120922066}.\footnote{Jack Dorsey, Vijaya Gadda, Tim Pool \& Joe Rogan.\\https://www.youtube.com/watch?v=DZCBRHOg3PQ}

\begin{table*}[]
\caption{Definition of classes and examples.}
\label{tab:class_defination}
\centering
\begin{tabular}{|l|l|l|}
\hline
\multicolumn{1}{|c|}{\textbf{Class name}} & \multicolumn{1}{c|}{\textbf{Definition of class}}                                                                                                                                                                                                    & \multicolumn{1}{c|}{\textbf{Examples}}                                                                       \\ \hline
Normal                                    & \begin{tabular}[c]{@{}l@{}}Any sample which does not troll, mock, insult \\ or threaten either an individual or a group.\end{tabular}                                                                                                                & \begin{tabular}[c]{@{}l@{}} \textit{``Coroner was a good career choice"} \end{tabular} \\ \hline
Profanity                                 & \begin{tabular}[c]{@{}l@{}}Samples that contain profane words that are not \\ directed towards a particular individual or group.\end{tabular}                                                                                                        & \begin{tabular}[c]{@{}l@{}} \textit{``What a fucking awful day"} \end{tabular}    \\ \hline
Trolling                                  & \begin{tabular}[c]{@{}l@{}}Content intended to cause disruption, trigger \\ conflict or insult for amusement.\end{tabular}                                                                                                                           & \begin{tabular}[c]{@{}l@{}} \textit{``Your body fat is as evenly distributed as the wealth} \\ \textit{in the US economy"} \end{tabular} \\ \hline
Derogatory                                & \begin{tabular}[c]{@{}l@{}}Insults and offensive content that is offensive and\\ directed to any group or individual, but does not \\ either constitute a direct threat or does not express \\ hatred towards that individual or group.\end{tabular} & \begin{tabular}[c]{@{}l@{}} \textit{``FUCKYOU U MATHRFUKER BITCH} \\\textit{IDIOOOOT NO BAN MEFROM EDIT I} \\\textit{TELL TRUTH}" \end{tabular} \\ \hline
Hate Speech                               & \begin{tabular}[c]{@{}l@{}}An expression of hatred towards individuals or \\ groups on the grounds of their identity.\end{tabular}                                                                                                                   & \begin{tabular}[c]{@{}l@{}}\textit{``Im going to start killing these assholes. Chin chin"}\end{tabular} \\ \hline
\end{tabular}
\end{table*}

The differences between trolling and hate speech can be subtle \cite{bjørkelo_2014} and often depend only upon degrees of sarcasm or aggression present in the text. Further, given the dichotomy between the relatively light-hearted nature of trolling and the extreme nature of hate speech, a gamut of sexist, racist, homophobic and otherwise offensive content exists that doesn't constitute extreme hate or direct threats. The social sciences have very well demarcated definitions for trolling, offensive content and hate speech \cite{hardaker2010trolling,bjørkelo_2014} which suggest that categories of offensive content can be demarcated based upon the severity or extremity of offence. 
To this end, the main contributions of this work are as follows:
\begin{itemize}
    \item To provide a publically available dataset\footnote{https://doi.org/10.5281/zenodo.3828501} of offensive content, with 12,490 samples, split across five classes - \textit{Normal}, \textit{Profanity}, \textit{Trolling}, \textit{Derogatory} and \textit{Hate Speech}. Table~\ref{tab:class_defination} relays a brief explanation of our classes which have been further elaborated upon in Section~\ref{class_definitions}. To our knowledge, this dataset is the first work that provides a manually labelled corpus that distinguishes between trolling and hate speech. 
    \item The data is sourced by reannotating samples from previously released public datasets \cite{founta2018large,davidson2017automated,holgate2018vulgar,gautam2019metooma} and thus contains tweets, comments from Wikipedia talk pages and Reddit posts. A diverse set of data sources reduces the potential presence of spurious statistical artefacts that could otherwise confuse and distract models. Section~\ref{data_sources} give more details about the distribution of samples from various platforms.
    \item To verify the absence of statistical artifacts via metrics introduced in \cite{niven-kao-2019-probing} (\textit{Applicability}, \textit{Productivity}, \textit{Coverage} and \textit{Strength}) and data ablation techniques \cite{NLPsCleverHansMomenthasArrived-2019-08-30}. 
\end{itemize}
BERT \cite{devlin2018bert} with default parameters achieves 75.3\% classification accuracy and 0.73 F1 score on our proposed dataset. The model also experiences an average drop of 0.27 on the F1 score when subjected to significant data ablation. This sensitivity indicates that our dataset is largely devoid of spurious statistical artefacts and can potentially lead to learning more robust models overall.

\section{Related Work}
\label{related_section}

The study of offensive content in social media broadly follows three major directions of inquiry - \textit{detection}, \textit{psychological implications} \cite{craker2016dark,suler2004online}, and \textit{human behaviour} \cite{buckels2014trolls}. Our work is mainly concerned with \textit{detection} that can be further categorized into three broad categories. \textit{the problem definition}, based on how precisely offensive content is defined; \textit{the granularity of classes}, indicating the different categories of offensive speech; and \textit{the feature modalities}, depending on the data modalities used as input features for classification models. Most early works in this domain are  characterized by broad, all-encompassing definitions of offensive content, binary categorization of offensive or not, and single modalities ---  either text, images, video, metadata, or networks. More recent work can be characterized by more precise definitions; fine-grained classification of hate speech categories and multiple modalities.

Early researchers formulated the problem as a binary classification task, with definitions of offensive content and its sub-categories being largely ambiguous. Terms with subtle distinctions like hate speech and cyberbullying were used interchangeably. Datasets were collected from social media platforms by searching for a limited number of handcrafted terms. Popular feature sets included token or character-level n-grams \cite{van2015detection}, sentence/document lengths \cite{dadvar2014experts}, capitalization \cite{nobata2016abusive,watanabe2018hate}, and document level sentiments \cite{chatzakou2017mean}. More recently, neural embeddings \cite{de2018modeling,ribeiro2018characterizing,zhang2018detecting}, have replaced the hand-engineered feature sets.

Colloquial and informal content produced in different social media channels pose challenges as discussed in Section~\ref{introduction_section}. To tackle these, researchers began using combinations of modalities. Studies in this vein started experimenting with user and post-level metadata along with textual features, age of the account \cite{al2016cybercrime}, the number of followers/followees \cite{zhong2016content}, presence of profanity in the username \cite{chatzakou2017mean}, and the presence of offensive terms in previous posts \cite{chen2012detecting,dadvar2013improving}. Researchers have also started looking at multimodal user-generated content \cite{zhong2016content,singh2017toward,hosseinmardi2015analyzing}. Dinakar et al. used knowledge graphs for detecting subtle and sarcastic trolling attempts \cite{Dinakar:2012:CSR:2362394.2362400}. Potha and Maragoudakis modelled the problem as a time series prediction problem \cite{potha2014cyberbullying}. Cheng et al. developed multimodal graph representations combining content, user data and metadata information \cite{cheng2019xbully}.

Recently, as meta-studies uncover major gaps in \textit{problem definition} and \textit{granularity of class} axes, research has started to expand in these directions \cite{Fortuna:2018:SAD:3236632.3232676,schmidt2017survey}. Davidson et al. proposed that abusive words can sometimes be used in a casual and inoffensive manner, different from hate speech \cite{davidson2017automated}. To this end, they proposed a dataset with three classes: \textit{Hate Speech}, \textit{Abusive} and \textit{Neither}. Davidson et al. also presented an extended version of the study discussing potential racial bias in offensive content datasets \cite{DBLP:journals/corr/abs-1905-12516}. Founta et al. proposed guidelines to create large offensive content datasets using crowdsourced workers \cite{founta2018large}, and \cite{malmasi2018challenges} conducted a set of experiments in a similar vein on a different dataset. Salminen et al. developed a highly granular taxonomy of different kinds of hate speech \cite{salminen2018anatomy}. It's relevant to mention that  \cite{de2018modeling} released an annotated corpus of trolling content, however the authors' intention was to attempt to model the poster's intentions and the affects on the recipient, and thus the messages are not guaranteed to contain trolling content. Further, the corpus is no longer publically available. Each of these studies demonstrated granularity of classes and precise definitions.

As the problems of hate speech and offensive content in social media grew in popularity - shared tasks and datasets began to be formed. The most notable of these are Track-1 of COLING 2018 \cite{kumar-etal-2018-benchmarking}, HatEval \cite{basile2019semeval} which sacrifices granularity in favour of multiple languages and OffensEval \cite{zampieri2019semeval} which attempts to identify whether offensive content is targeted or not. These datasets also pose the detection of offensive content as a binary classification problem. 

Our work explores granular classes from a different angle, distinguishing between offensive content based upon the severity of offense. Our dataset is best categorized as having precise definitions for offensive content, having moderate granularity and using a singular modality of textual content.

\section{Data Sources }
\label{data_sources}

To annotate offensive content, annotators typically need to winnow through a large number of innocuous samples to find a significant amount of offensive content. To this end, we collect data by relabelling randomly selected samples from publicly available datasets; \cite{davidson2017automated}, \cite{founta2018large}, \emph{The Kaggle Jigsaw Toxic Comments}\footnote{https://www.kaggle.com/c/jigsaw-toxic-comment-classification-challenge}, \cite{holgate2018vulgar} and \cite{gautam2019metooma}. For the rest of this work, we refer to them as  \emph{Davidson}, \emph{Crowdsource}, \emph{Jigsaw}, \emph{Why Swear} and \emph{\#MeToo}, respectively. 

Davidson collected a set of 24,000 tweets and labelled them as hate speech, offensive language or neither. Their definition of hate speech is quite broad relative to ours, as they include all language that is intended to humiliate or derogatory towards an individual or group. They further report that classifiers as well as human annotators tend to confuse their hate speech and offensive language classes. They conclude that future work needs to consider social context in the task of hate speech detection and ensure that hate speech categories do not only contain multiple extreme slurs. 

Crowdsource is a set of 100,000 tweets labelled as abusive, normal, hateful and spam. The authors use a similar definition of hate speech as Davidson. However they find that an annotation scheme that distinguishes between sub categories of abusive and hateful content is unsuitable for large scale crowd annotation. 

Jigsaw is a set of 300,000 comments from Wikipedia Talk Pages that are annotated for six classes: toxic, severe toxic, insult, obscene, threat and identity hate. It was originally used for a toxic comment detection challenge on Kaggle.

Davidson, Crowdsource and Jigsaw are three very popular datasets that deal with hate speech and offensive cotent and are thus good potential sources of data. Why Swear seeks to analyse the role that vulgar words play in the detection of offensive content. Since previous work has shown that the task of offensive content detection can be biased towards the presence of offensive words, using Why Swear as a source corpus meant a reasonable chance of having samples/phrases with vulgar words used in non-offensive context and potentially, a less biased dataset. 

MeToo represents a corpus of text that has been annotated for hate speech, relevance and sarcasm - which are important factors in our own annotation system. Further, the dialogue or content in these samples is likely to be focused around sexism which may not have been reflected in the Jigsaw data. 

\begin{figure}[!htbp]
\includegraphics[width=\columnwidth]{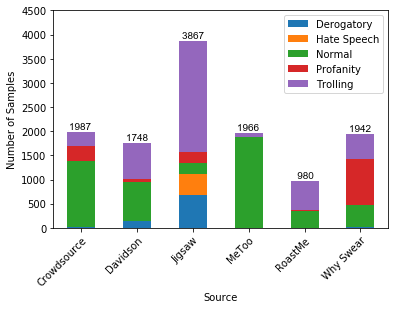}
\caption{Distribution of classes by source. \#MeToo is overwhelmingly Normal, and Jigsaw contributes the lion's share of Trolling and Hate Speech.}
\label{fig:dist_class_source}
\end{figure}

In addition to sampling previously related datasets, we augment the data with a thousand samples from the RoastMe Subreddit\footnote{https://www.reddit.com/r/RoastMe/} as the forum tends to focus insults on appearance which might have been underrepresented in other platforms. Collectively, our data  contains samples from Twitter, Talk Pages of Wikipedia articles and Reddit. Figure~\ref{fig:dist_class_source} shows the distributions of annotated classes from each source dataset.

\section{Class Definitions}
\label{class_definitions}
Offensive content runs the gamut between casual pejoratives, slurs, and threats of violence. Combining the different kinds of offensive content into a single bucket is thus inappropriate. Our dataset attempts to differentiate based upon the degree of offence by labelling samples into five classes: \textit{Normal}, \textit{Profanity}, \textit{Trolling}, \textit{Derogatory} and  \textit{Hate Speech} with each of them defined as:
\begin{itemize}
        \item \textit{Normal}: Any sample which does not troll, mock, insult or threaten either an individual or a group. Examples are \textit{``Coroner was a good career choice"} and \textit{``The persecution of gay people must be stopped"}.
        \item \textit{Profanity}: Content that contains profane words that are not directed towards a particular individual or group. For example, \textit{``What a fucking awful day"}.
        \item \textit{Trolling}: Content intended to cause disruption, trigger conflict or insult for amusement. Users who participate or conduct trolling are called trolls \cite{hardaker2010trolling}. For example, \emph{``You look like the generic gay hipster that has too high of an ego"}.
        \item \textit{Derogatory}: Insults and messages that are offensive and directed to any group or individual, but do not constitute a direct threat or express hatred towards that individual or group, \emph{e.g.}, \textit{``O FUCK YOU U MATHRFUKER BITCH IDIOOOOT NO BAN ME FROM EDIT I TELL TRUTH"}.
        \item \textit{Hate Speech}: An expression of hatred towards individuals or groups on the grounds of their identity - political stance, religious belief, race and ethnicity, national origin or sexual orientation \cite{bjørkelo_2014}. Examples are \textit{``you accuse me of vandalism, i'll vandalize yo face, nigga"} and \textit{``I'm going to start killing these assholes. Chin chin."}
\end{itemize}
A pertinent point of distinction between hate speech and trolling is the presence of viciousness or aggression. Hate speech samples are significantly violent or extremely offensive. For example, the comparison of the phrases \textit{``you're so gay"} and \textit{``every gay boy deserves to be slaughtered"} reveals the latter to be significantly more vicious. Thus the former is considered trolling and the latter, hate speech despite their both using the word ``gay" in the same context. Similarly, the point of distinction between derogatory, trolling and hate speech is in the aggression displayed or the terminology used. The phrase \textit{``Stop acting like a fag"} is too offensive to be trolling because of the word choice. Despite being homophobic, it does not necessarily express hatred towards homosexuals and thus belongs in the derogatory class. On the other hand, phrases like \textit{``all gults should get cancer"} is classified as hate speech even though the phrase ``gult" is a relatively inoffensive slur.

\section{Annotation Process and Guidelines}
\label{annotation_guidelines}

The data was annotated independently by a PhD student and two research assistants, each of whom were intimately familiar with this domain. They were instructed to be wary of seemingly innocuous samples that could contain words or phrases considered offensive in other cultures. Further, they were advised to make a decision keeping the entire sample in mind, rather than the presence of highly offensive words. In the initial draw, a random set of 2,000 samples was picked from each parent dataset\footnote{including 1000 samples from RoastMe} and annotated. Post a preliminary stage of annotation, a second set of 2,000 samples was picked from Jigsaw's  Obscene, Toxic and Severely Toxic classes, in order to boost derogatory and hate speech samples. Once all the samples were annotated, duplicate rows, blank rows and samples containing other languages were removed, resulting in slightly less than 13,000 samples. Of these, the annotators could not agree on 495 samples which were also removed, making the final dataset size 12,490 samples. The annotation guidelines are as follows.

\subsection{Profanity Detection}
Content with the presence of vulgarity, profanity or swear/curse which is not directed to an individual or a group. The following are examples of profanity:
\begin{enumerate}
\item \emph{``What the fuck is wrong with this day? Can it get any worse?"} - The profane word ``fuck" is not directed as an insult towards any individual or group. Thus, the profanity class is appropriate.
\item The phrase ``my nigga" in \emph{``This ghetto is full of my nigga"}, is loosely equivalent to ``my friends". The context is somewhat ambiguous, but the word ``nigga" is not being used in a derogatory fashion or victimizing a particular individual or group. So, profanity is appropriate here. 
\end{enumerate}
\subsection{Trolling Detection}
Content intended to cause disruption,  trigger conflict or mild insults for amusement. It may have a sense of humour, sarcasm or mockery. It can be directed to a group of people or an individual. The following are the examples of trolling:
\begin{enumerate}
    \item \emph{``Your body fat is as evenly distributed as the wealth in the US economy"}. This message contains no profane words, mocks an individual and is not an expression of hatred or extreme insult. Thus trolling.
    \item \emph{``Fuck you Fuck you Fuck you Fuck you ... Fuck you''} The word ``Fuck" is directed towards a particular individual or group; thus it cannot be profanity. Continuous repetition, without any further context, suggests that the message is intended to be disruptive, rather than express hatred or insult. Thus the hate speech and derogatory classes are inappropriate, and the message should be labelled trolling.
    \item \emph{``3 MOST COMMON WHITE PEOPLE ACTIVITIES: 1. Wearing Tapout, 2. Shooting up schools,3. Calling you a nigger on Xbox-"} The message quite obviously is directed towards white people; thus profanity is not suitable. The very offensive word ``nigger" makes the derogatory or hate speech classes very tempting, however, since the overall message is intended as a mockery of white people, trolling is the most appropriate.
    \item \emph{``My grandma used to call me a porch monkey all the time... Come to think of it she did refer to a broken bottle as a nigger knife."} This sample contains a large number of racist remarks and is quite offensive. However, in this particular case, the comment is a reference to a scene from a movie that is a somewhat popular meme, and thus this message is most likely trolling.
\end{enumerate}
\subsection{Derogatory Content Detection}
A harsh, serious insult that is hurtful and derogatory to an individual or a community. These are the comments which are too severe to be considered trolling and yet do not express hate. The following are examples of offensive content:
\begin{enumerate}
    \item \emph{``All muslims are gay"}. In this message, even though the word ``gay" is not profane or offensive in of itself, it is being used as a pejorative for Muslims; thus, the derogatory class is appropriate.
    \item \emph{``Guys are not supposed to get their ass sucked queer!"} In context, the otherwise innocuous word ``queer" is assumed to be an insult. Despite an attempt at juvenile humour, the message is quite offensive and belongs in the derogatory class.
\end{enumerate}
\subsection{Hate Speech Detection}
Extremely violent or offensive content, as well as identity-based hate. Threats of violence towards a person or group of people based on their religious beliefs, ethnicity, race, disability, gender, sexual orientation or political inclination. Some examples of hate speech are as follows:
\begin{enumerate}
    \item \emph{``White Trash Fuck off you white piece of trash. Fucking wiggers. Go and fuck a Jew, Jew raper."} This sample is offensive and expresses hatred towards white people. Quite simply, hate speech. 
    \item \emph{``I'll kick the shit out of you ya cunt"} This sample is a threat of violence, potentially to a woman. Undoubtedly hate speech. 
    \item \emph{``WHO CARES ABOUT A BUNCH OF DIRTY MONKEYS FROM HAITI. MONCRIEF, WHAT ARE YOU A JEW?."} This sample expresses hatred for people from Haiti and tries to imply that they are sub-human. For this reason, hate speech.
\end{enumerate}

\begin{figure}[!htbp]
\includegraphics[width=\columnwidth]{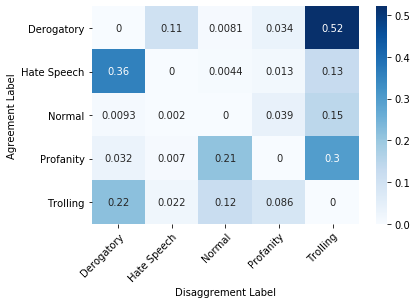} 
\caption{Confusion matrix between majority agreement label and minority annotator. 52\% of the derogatory samples had one out of three annotators believe it should be trolling.}
\label{fig:confusing_classes}
\end{figure}

The annotator agreement between three annotators was measured by the \textbf{Randolph (0.65)} and \textbf{Fleiss Kappa (0.59)}. Both these scores are indicative of Moderate Agreement. Fig.~\ref{fig:confusing_classes} shows the most confused classes between the three annotators. The class with majority agreement is plotted on the Y-axis, and the third annotator's decision is on the X-axis. We observe that trolling and derogatory classes are most frequently confused; of the derogatory class, 50\% of the samples have a disagreeing annotator believe that they should have been marked trolling. Similarly, for the hate speech class, roughly 36\% of the samples had a dissenting annotator believe that they should have been marked as derogatory. We believe this reflects the subjective nature of offensive content. The confusion between trolling and profanity is incongruous, as the class definitions are quite distinct.

% \begin{table}[]
% \caption{Inter Annotator Agreement Score for Samples by each parent dataset.}
% \label{tab:Source_wise_inter_annotator_score}
% \centering
% \begin{tabular}{|l|l|l|}
% \hline
% \textbf{Source}   & \textbf{Fleiss Kappa} & \textbf{Randolph kappa} \\ \hline
% Subreddit RoastMe & 0.418                 & 0.6333                  \\ \hline
% Crowdsource       & 0.746                 & 0.840                   \\ \hline
% Jigsaw            & 0.481                 & 0.599                   \\ \hline
% Why Swear         & 0.518                 & 0.597                   \\ \hline
% \#MeeToo          & 0.201                 & 0.828                   \\ \hline
% T.Davidson        & 0.349                 & 0.444                   \\ \hline
% \end{tabular}
% \end{table}

\section{Dataset and Metadata Analysis}
\label{dataset}

\begin{figure*}[!htbp]
\centering
\includegraphics[width=\textwidth]{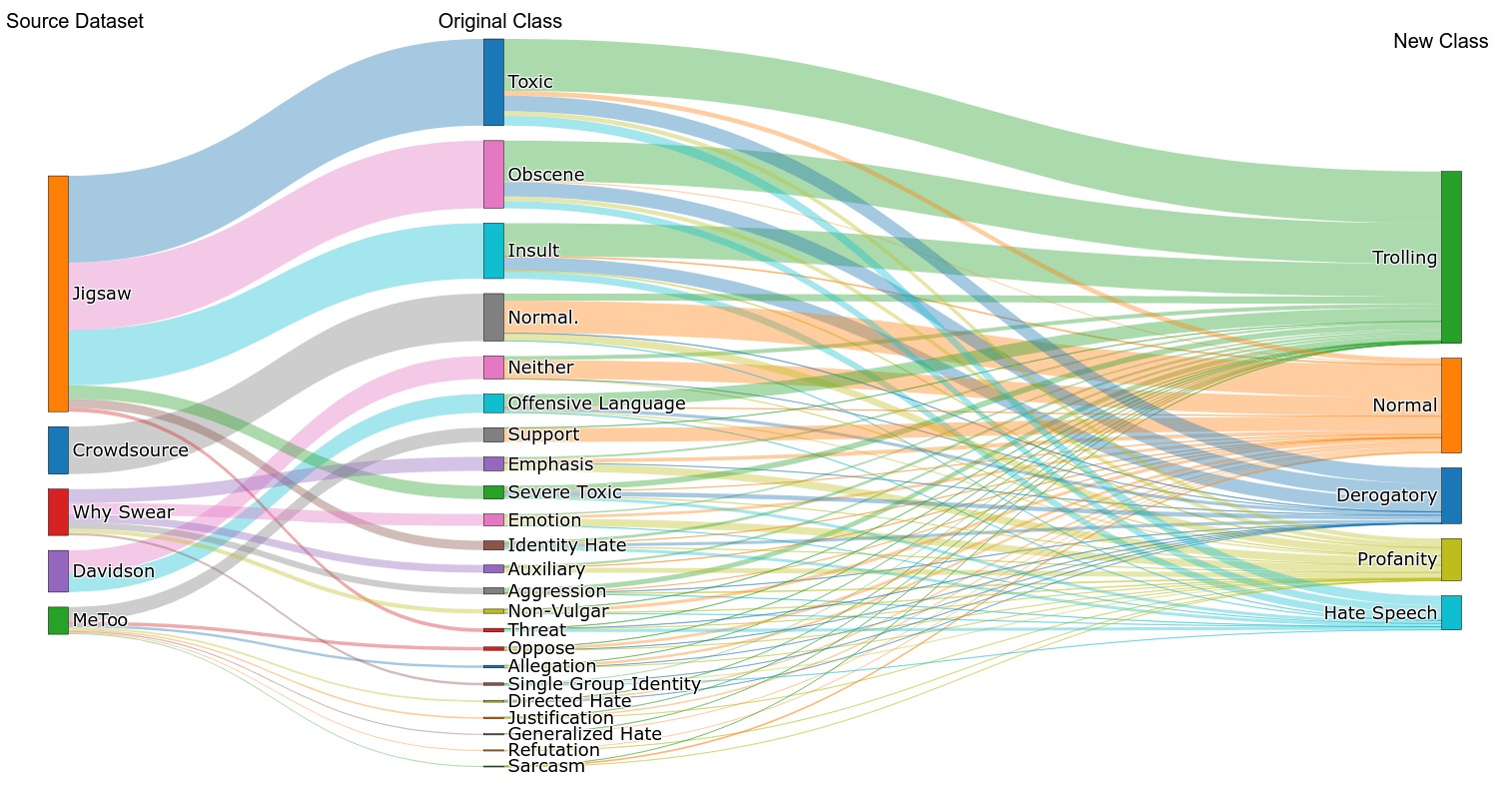}
\caption{Class change from source datasets to ours. Jigsaw and MeToo were originally multi-label which skews this figure slightly.}
\label{fig:change_in_class}
\end{figure*}

\begin{figure}[!htbp]
\centering
\scalebox{0.95}{
\includegraphics[width=\columnwidth]{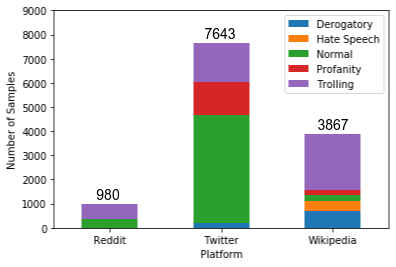}}
\caption{Class distribution by platform. Twitter contributes the most samples overall, though most trolling comes from Wikipedia.}
\label{fig:dist_class_platform}
\end{figure}

\begin{table}[]
\caption{Class distribution in our Dataset. Severity of offensive and number of samples are inversely proportional.}
\label{tab:Class_stats}
\centering
\begin{tabular}{|l|l|}
\hline
\textbf{Class} & \multicolumn{1}{c|}{\textbf{\begin{tabular}[c]{@{}c@{}}\# of Samples\end{tabular}}} \\ \hline
Normal         & 5,053                                                                                       \\ \hline
Trolling       & 4,537                                                                                       \\ \hline
Profanity      & 1,582                                                                                       \\ \hline
Derogatory     & 862                                                                                        \\ \hline
Hate Speech    & 456                                                                                        \\ \hline
\end{tabular}
\end{table}
The dataset has 12,490 Samples, split across five classes. The distribution of classes in the dataset is in Table~\ref{tab:Class_stats}. Figure~\ref{fig:dist_class_source} lists the distribution of samples mined from each source. These sources include data from different social media platforms. Figure~\ref{fig:change_in_class} shows the change of the samples from each original class in their respective datasets and new actual annotation classes. A point of interest is that \textbf{samples from virtually every label of the source datasets have been mapped to trolling.}
Figure~\ref{fig:dist_class_platform} lists the distribution of samples from each platform. Despite the large majority of the data coming from Twitter, a third of the total samples come from Wikipedia Talk Pages. 

Collectively, we observe a decrease in the number of samples as the severity of offensive content increases. This is to be expected. A large proportion of samples taken from Crowdsource are assigned the normal class. Since Crowdsource is imbalanced and close to 60\% of the source data is Normal, this is also expected. Surprisingly, MeeToo samples are also disproportionately biased towards the normal class. On inspection, the majority of MeeToo samples are either accusations of harassment or discussion around various allegations and do not constitute threats or insults which explains this discrepancy. Jigsaw supplies a large amount of the hate speech in our dataset, largely because of the second round of annotations that was carried out only with samples from Jigsaw's more offensive classes. Finally, Why Swear provides the lion's share of the profanity class. 

To help identify edge cases and points of failure, we examine the distribution of the type of offensive content present in the data. We use the vocabulary from Hatebase.org\footnote{https://hatebase.org/} to assign each sample to one or more categories. The percentages of each category are presented in Table~\ref{tab:abusive_term_class}. Samples of hate speech present in our dataset are dominated by insults directed at people's gender and sexual orientation. The fact that hate speech has a more significant proportion of these slurs is to be expected - though this necessitates examining these terms as potential cues for models trained on our data to learn instead of learning from the overall context of the samples.

\begin{table}[]
\caption{Percentage of vulgar terms type by label. Gender and Orientation are represented more than other types.}
\label{tab:abusive_term_class}
\centering
\scalebox{0.75}{
\begin{tabular}{|l|l|l|l|l|l|}
\hline
\textbf{Abuse Type} & \textbf{Normal} & \textbf{Profanity} & \textbf{Trolling} & \textbf{Derogatory} & \multicolumn{1}{c|}{\textbf{\begin{tabular}[c]{@{}c@{}}Hate \\ Speech\end{tabular}}} \\ \hline
\textbf{Orientation} & 0.3             & 0.70      & 2.49     & 15.31               & 15.79                \\ \hline
\textbf{Gender}      & 2.97            & 8.91      & 19.97    & 35.85               & 20.17                \\ \hline
\textbf{Disability}  & 0.40            & 0.44      & 4.74     & 4.88                & 5.48                 \\ \hline
\textbf{Ethnicity}   & 8.97            & 3.22      & 5.64     & 8.93                & 10.31                \\ \hline
\textbf{Nationality} & 4.06            & 0.50      & 0.66     & 0.93                & 1.53                 \\ \hline
\textbf{Religion}    & 0.91            & 0.06      & 0.13     & 0.12                & 1.09                 \\ \hline
\textbf{Class}       & 0.43            & 0.06      & 0.26     & 0.12                & 0.0                  \\ \hline
\end{tabular}}
\end{table}

\begin{figure*}[!htbp]
\centering

\begin{minipage}{0.32\textwidth} 
\includegraphics[width=\linewidth]{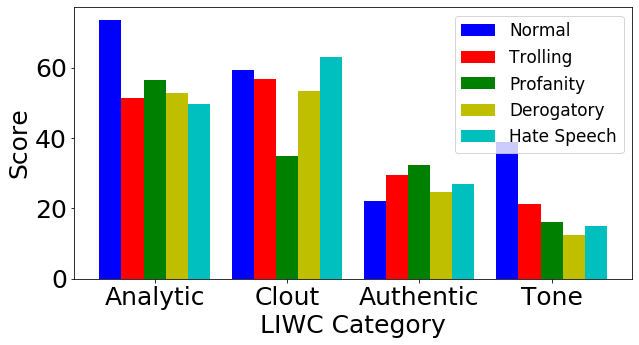}    
\subcaption{Summary}    
\end{minipage}    
\hspace{\fill}
\begin{minipage}{0.32\textwidth} 
\includegraphics[width=\linewidth]{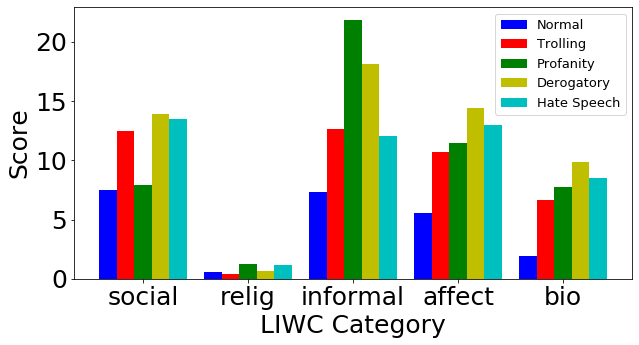}    
\subcaption{Psychological Processes}    
\end{minipage}  
\hspace{\fill}
\begin{minipage}{0.32\textwidth} 
\includegraphics[width=\linewidth]{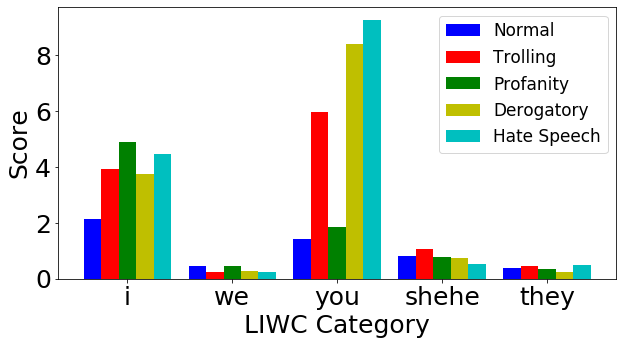}    
\subcaption{Person Pronouns}    
\end{minipage} 

\vspace{0.1cm}

\begin{minipage}{0.32\textwidth} 
\includegraphics[width=\linewidth]{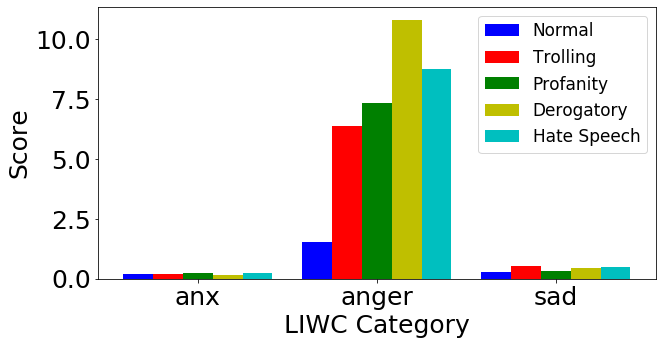}    
\subcaption{Negative Emotions}    
\end{minipage}    
\hspace{\fill}
\begin{minipage}{0.32\textwidth} 
\includegraphics[width=\linewidth]{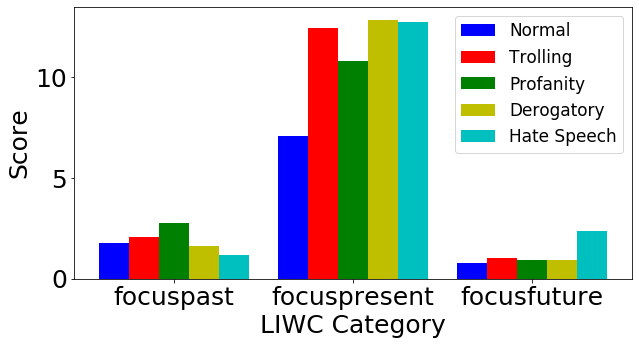}
\subcaption{Temporal Focus}    
\end{minipage}  
\hspace{\fill}
\begin{minipage}{0.32\textwidth} 
\includegraphics[width=\linewidth]{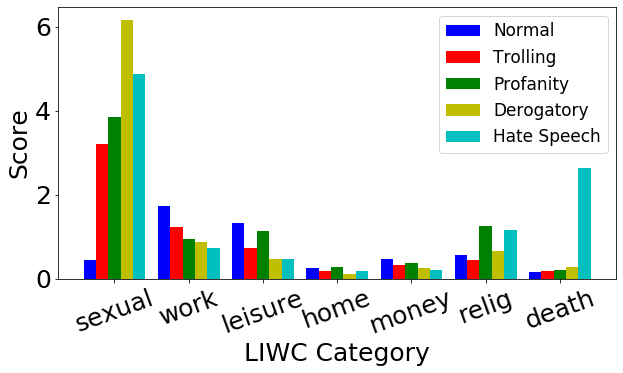}
\subcaption{Personal Concerns}    
\end{minipage}

\caption{Mean scores for LIWC categories. The data is dominated by samples that use present tense, and display anger, as expected of such a corpus. characteristic differences exist between classes which are in line with our definitions. For example, (c) derogatory focuses on the pronoun ``you" but profanity uses ``I". (f) hate speech displays a high percentage of death content (expected due to it's extreme and hurtful nature)}
\label{fig:LIWC}
\end{figure*}

Finally, Figure~\ref{fig:LIWC} shows the mean scores for LIWC categories across all the classes of our dataset. Much has already been said in literature about LIWC categories \cite{elsherief2018hate,pennebaker2015development} However, a few interesting observations emerge in the analysis. The Personal Concern Death is dominated by the hate speech class, which is due to direct threats being categorized as hate speech. The high representation of the sexual personal concern across most classes is likely due to the frequency of vulgar words like ``fuck" across 4 of the 5 categories. A somewhat unexpected outcome is that derogatory has a greater percentage of anger than hate speech since hate speech should contain more anger or be more extreme in terms of offense or anger. The profanity class has ``I" as the most common personal pronoun, by a small margin. The personal pronoun ``you" is a lot more common in hate speech and derogatory because of directed insult and abuse in those classes.

\section{Dataset Quality Analysis}
\label{analysis}
Studies on the composition of text datasets have shown that state-of-the-art models can ``short" and fit on distributions of tokens rather than gain some understanding of language \cite{niven-kao-2019-probing}. Models trained on datasets with a large proportion of very effective cues could potentially learn to associate cues with specific labels and disregard any linguistic information. We evaluate our dataset on these metrics to determine potential cues. Dataset ablations help verify their accuracy.

\subsection{Classification Models}
\noindent \textbf{Data Preprocessing} - Text is preprocessed using the fast.ai Tokenizer\footnote{https://docs.fast.ai/text.transform.html\#Tokenization} and follows its conventions. Token ``xxup'' is inserted right before capitalised words, token ``xxmaj" is inserted before words in title case. Repeated words and characters are further removed after placing appropriate marking tokens. The data is anonymised - any personally identifiable information like Twitter username and IP address are removed, and hashtags are converted into regular words. We remove special characters and XML Tags and also, replace URLs with the token ``xxurl". Tokenisation for BERT is done using BERT fulltokenizer\footnote{https://github.com/google-research/bert/blob/master/\\tokenization.py}. The samples are then divided into an 80-20 training and validation split. 

Three different models, a character CNN \cite{DBLP:journals/corr/ZhangZL15}, a sentence-level CNN \cite{kim2014convolutional} with Fasttext embeddings and BERT-base-cased \cite{devlin2018bert} are trained. The best performing model is selected for the calculation of \textit{Productivity} and tested for sensitivity to data ablation. No major hyperparameter optimization is done, and yet all three models perform very well out of the box. Table~\ref{tab:model_performance} shows the model performance on the validation set.

\begin{table}[]
\caption{Model performance on our dataset.}
\label{tab:model_performance}
\centering
\begin{tabular}{|l|l|l|}
\hline
\textbf{Model} & \textbf{\begin{tabular}[c]{@{}l@{}}Accuracy\end{tabular}} & \multicolumn{1}{c|}{\textbf{\begin{tabular}[c]{@{}c@{}}Weighted F1 \\ Score\end{tabular}}} \\ \hline
Naive Bayes     & 64.4\%                                                                  & 0.57                                                                                       \\ \hline
Character CNN  & 68.3\%                                                                  & 0.65                                                                                       \\ \hline
N-gram CNN     & 70.4\%                                                                  & 0.68                                                                                       \\ \hline
BERT           & \textbf{75.3\%}                                                    &\textbf{0.73}                                                                                        \\ \hline
\end{tabular}
\end{table}

The focus of this work is not on the classifiers, however for an in-depth view of all preprocessing steps and classifier hyperparameters (see the attached code).\footnote{Inserted after acceptance.}

\subsection{Post Length Cue}
\begin{figure}  
\centering
\begin{minipage}{0.45\columnwidth} 
\includegraphics[width=\linewidth]{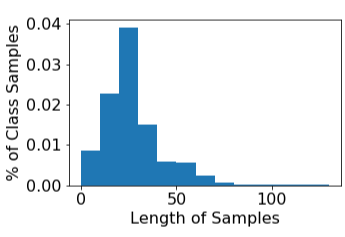}    
\subcaption{Normal}    
\end{minipage}
% \hspace{1cm}
\begin{minipage}{0.45\columnwidth} 
\includegraphics[width=\linewidth]{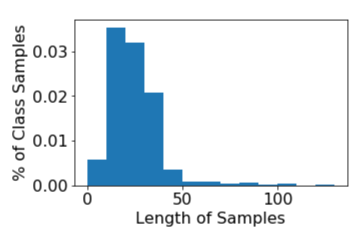}    
\subcaption{Profanity}    
\end{minipage}

\vspace{0.1cm}

\begin{minipage}{0.45\columnwidth} 
\includegraphics[width=\linewidth]{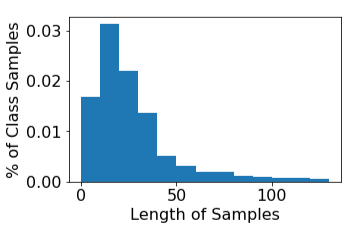}    
\subcaption{Trolling}    
\end{minipage}
\begin{minipage}{0.45\columnwidth} 
\includegraphics[width=\linewidth]{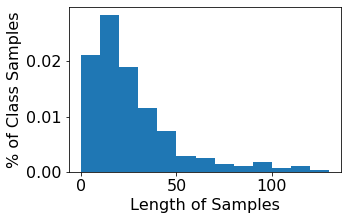}    
\subcaption{Derogatory}    
\end{minipage}

\vspace{0.1cm}
\begin{minipage}{0.45\columnwidth} 
\centering
\includegraphics[width=\linewidth]{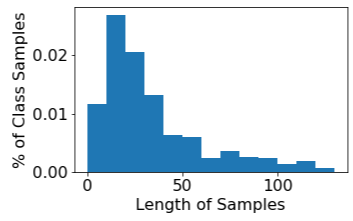}    
\subcaption{Hate Speech}    
\end{minipage}

\caption{Sample length distribution for each class. Majority of distribution for all classes follows a similar distribution; this reduces the potential for a model to associate a label with sample length.}
\label{fig:length_dist_class}

\end{figure}

One of the benefits of the data being sourced from multiple platforms is the potential for models trained on the data to learn how to detect offensive content in a more generalised fashion. However, a disadvantage is that each platform has a different idiomatic style of messages that is unique to that platform. For example, conversations on Twitter tend to have short messages, with threads being used to convey long-form ideas. Thus the length distribution of samples across labels is important - otherwise, models could potentially treat the length of the input sample as a cue for that class. Figure~\ref{fig:length_dist_class} shows the length distribution of each class and shows that the class lengths are similarly distributed, with very few samples in normal, trolling and hate speech being significantly longer.

\subsection{N-gram Cues}
Unigram, bigram, and trigram sized tokens of each sample are extracted and evaluated based on the following metrics.
    
\noindent \textbf{Applicability}: Given a token unique to a single class, the \textit{Applicability} of the token is the count of data samples which contain that token. Mathematically, the Applicability for a token $k$ is defined in Eq.~\ref{app_eq}. Intuitively, this metric speaks to potential cues for the model to associate with that class. 
    
\begin{equation}
    \alpha _{k} =\sum_{i=1}^{n} \mathbbm{1}\left [ \exists j,k \in \mathbbm{T}_{j}^{(i)} \wedge k\notin \mathbbm{T}_{\neg j}^{(i)} \right ]    
    \label{app_eq}
\end{equation}
    
Here, $\mathbbm{T}_{j}^{(i)}$ is set of all tokens in subset for data point $i$ with label $j$. 
    
\noindent \textbf{Productivity}: Given a token unique to a single class, the \textit{Productivity} of the token is the proportion of data samples for which the model predicts the correct answer, relative to the \textit{Applicability} of the token. Therefore, the \textit{Productivity} of a token is a metric for how useful the model would find it to be. \textit{Productivity} $\pi_{k}$ can be calculated as defined in Eq.~\ref{pro_eq}.
    
\begin{equation}
    \pi_{k} =\frac{\sum_{i=1}^{n} \mathbbm{1}\left [ \exists j,k \in \mathbbm{T}_{j}^{(i)} \wedge k\notin \mathbbm{T}_{\neg j}^{(i)} \wedge y_{i}=j \right ]}{\alpha_{k}} 
    \label{pro_eq}
\end{equation}

\noindent \textbf{Coverage} - Coverage is simply the proportion of \textit{Applicability} relative to the total number of rows in the dataset.

\begin{equation}
    C_{k} = \frac{\alpha _{k}}{n}
    \label{cov_eq}
\end{equation}
    
\noindent \textbf{Strength} - The strength of a cue is defined as the product of its \textit{Coverage} and \textit{Productivity}.
\begin{equation}
    S_{k} = C_{k} \times \pi_{k}
    \label{st_eq}
\end{equation}

We extract cues from our dataset and calculate its \textit{Productivity}, \textit{Coverage} and \textit{Strength}. Cues with non zero Applicability (Eq.~\ref{app_eq}) or Coverage (Eq.~\ref{cov_eq}) primarily serve as characteristic cues for each label. For all three datasets, the large proportion of cues show an inverse relationship between \textit{Productivity} and \textit{Coverage}, this would imply that most cues are either widespread (high \textit{Applicability}) or very profitable for the model to learn (high \textit{Productivity}). Further, the Applicability of the strongest cues is quite low - this means that the majority of these cues are cues because they occur in single-digit samples of one class throughout the dataset. 
    
Table~\ref{tab:cues} shows the top 5 cues for both the training and validation sets. Both, the strongest cues for the normal, trolling and profanity classes seem to be fairly random. This is a little surprising since intuitively one would expect the profanity class to be characterised by profane words. However, given the nature of profane words, it is unlikely for insults to be restricted to profanity, without also showing up in either derogatory or hate speech, if not both. It stands to reason that unigrams will generally be significantly stronger than bigrams or trigrams since unigrams will occur more frequently, increasing their \textit{Coverage} and \textit{Productivity}. Cues for derogatory and hate speech definitely include profane words, and this is a cause for concern. However, the cues for these classes are not unigram cues. bigram or trigram cues will have lower Applicability and Coverage as it is less likely for longer sequences to repeat themselves in the corpus. 
\subsection{Data Ablation}

Cues across labels can distract models from learning context or meaningful information from data.
We perform data ablation tests proposed by \cite{NLPsCleverHansMomenthasArrived-2019-08-30}. In principle, models should be susceptible to significant dataset ablations - as the dataset is transformed dramatically, model performance should drop precipitously.
    
\noindent \textbf{Scramble Word Order}: The tokens or words for each sample in the test set are randomly shuffled. A drop in test set accuracy indicates that the model depends upon the sequential nature of words/cues to at least some extent. If the test set performance does not change significantly, this could indicate that the model has effectively learned a bag-of-words style classification.
    
\noindent \textbf{Shuffle Labels}: The class labels are randomly shuffled, and the model is retrained. If test set performance does not change significantly, this indicates that the model has not learned to associate contextual cues with each label. This is the only ablation test that involves model retraining instead of perturbing inputs of the previously trained model.
    
\noindent \textbf{Partial Input}: Tokens from the samples are removed with a probability of 0.5, but the sequential order of the remaining tokens is not disturbed. If test set performance does not drop, this indicates that the model depends upon a subset of tokens within each sample to make a prediction.
    
In general, for a model to be learning linguistic information from the dataset, we would expect: (i) data should contain fewer cues of considerable strength and (ii) model performance should drop appreciably with each ablation. Table~\ref{tab:char_ann_ablation_study} shows the drops in performance for each ablation process. the sensitivity of the models to each ablation suggests that the model depends on the entire message to make predictions rather than shorting on statistical artefacts.
% \if 0
% \begin{itemize}
%     \item The data should contain fewer cues of considerable strength.
%     \item Model performance should drop appreciably with each ablation.
% \end{itemize}
% \fi 

\begin{table}[]
\small
\caption{BERT ablation study results.}
\label{tab:char_ann_ablation_study}
\centering
\begin{tabular}{|l|l|l|}
\hline
\textbf{Ablation Technique} & \textbf{\begin{tabular}[c]{@{}l@{}}Accuracy\end{tabular}} & \multicolumn{1}{c|}{\textbf{\begin{tabular}[c]{@{}c@{}}Weighted F1 \\ Score\end{tabular}}} \\ \hline
No Ablation                 & \multicolumn{1}{c|}{75.3\%}                                             & \multicolumn{1}{c|}{0.73}                                                                  \\ \hline
Shuffling Labels            & 40.2\%  (-35.1\%)                                                           & 0.23 (-0.50)                                                                               \\ \hline
Scrambling Words            & 62.4\% (-12.8\%)                                                            & 0.58 (-0.15)                                                                               \\ \hline
Removing Words              & 60.8\% (-14.4\%)                                                            & 0.56 (-0.17)                                                                               \\ \hline
\end{tabular}
\end{table}

\begin{table}[]
\caption{The top 5 cues by strength for each class in the proposed dataset. The absence of strong unigram cues for all classes is a positive sign.}
\label{tab:cues}
\centering
\scalebox{0.8}{
\begin{tabular}{|l|l|l|}
\hline
\textbf{Data Partation}         & \textbf{Class} & \textbf{Top 10 Cues}                                                                                                                                          \\ \hline
\multirow{5}{*}{Training Set}   & Normal         & \begin{tabular}[c]{@{}l@{}}diagnosed, getup deep dish, \\ dish vocal mix, yellow asian flu, \\ plated letz getup\end{tabular}                                 \\ \cline{2-3} 
                                & Trolling       & \begin{tabular}[c]{@{}l@{}}scumbag Ban, de Barranquilla OMG, \\ barranquillawtf come Junior, \\ push revert nt, maybe Junior de\end{tabular}                  \\ \cline{2-3} 
                                & Profanity      & \begin{tabular}[c]{@{}l@{}}nt exactly, spiritual, know hell, \\ Stg, ass b number\end{tabular}                                                                \\ \cline{2-3} 
                                & Derogatory     & \begin{tabular}[c]{@{}l@{}}eternityfaggot, little better m, \\ love doggy style, letting smell like,\\ vagina Insist letting\end{tabular}                     \\ \cline{2-3} 
                                & Hate Speech    & \begin{tabular}[c]{@{}l@{}}ALIVE, die soon, \\ TRUTH u GODDAMN, \\ FUCK RACIST CANUCK,\\ ANUCK PIECE o\end{tabular}                                           \\ \hline
\multirow{5}{*}{Validation Set} & Normal         & \begin{tabular}[c]{@{}l@{}}programme, work NICU tchhr, \\ Yes surprisedcongrats friends, \\ friends work NICU, \\ surprisedcongrats friends work\end{tabular} \\ \cline{2-3} 
                                & Trolling       & \begin{tabular}[c]{@{}l@{}}create elearnign, defiencent assume d, \\ single fucking Hawiye, Let tell thing,\\ said radiofan patience\end{tabular}             \\ \cline{2-3} 
                                & Profanity      & \begin{tabular}[c]{@{}l@{}}eno, Gon na piss, find shirt want, \\ know wo nt, na piss b\end{tabular}                                                           \\ \cline{2-3} 
                                & Derogatory     & \begin{tabular}[c]{@{}l@{}}HACK, total fucking faggot, \\ Congrats total fucking, SHUT FAG \\ LOL, HUHOOH M SHAKIN\end{tabular}                               \\ \cline{2-3} 
                                & Hate Speech    & \begin{tabular}[c]{@{}l@{}}YOUGO KILL, tranny surgery tit, \\ Kim Kardashian vandals, \\ end jizz shoter, pissing seat tell\end{tabular}                      \\ \hline
\end{tabular}}
\end{table}

\section{Conclusion and Future Work}
\label{conclusion}
In conclusion, trolling is a sub-category of offensive content, that can frequently be mislabelled as either severely offensive content or hate speech. In order to prevent overzealous censorship in the name of hate speech removal, it's important to acknowledge the existence of trolling. To the best of our knowledge, this dataset is the first work that distinguishes between trolling and hate speech in the context of offensive content detection. The dataset will be publically available via Zenodo (\textit{https://doi.org/10.5281/zenodo.3828501}) and encompasses samples from multiple platforms in social media. Care has been taken to ensure the dataset is largely devoid of spurious statistical artefacts and is shown to result in models that learn sequential cues and potentially linguistic information instead of just bags of words. The data validation and analysis metrics could also be used in the future as an alternative method for evaluating datasets in this domain.

\fontsize{9.0pt}{9.0pt}
\bibliography{ref.bib}
\bibliographystyle{aaai}

\end{document}